\newcommand{\rem}[1]{}
\documentclass{amsart}
\usepackage{amsfonts,amssymb,amsmath,amsthm,mathrsfs}
\usepackage{url}
\usepackage[dvips]{epsfig}
\newtheorem{thrm}{Theorem}[section]
\newtheorem{lem}[thrm]{Lemma}
\newtheorem{prop}[thrm]{Proposition}

\theoremstyle{definition}

\begin{document}
\author[C.~A.~Mantica and L.~G.~Molinari]
{Carlo~Alberto~Mantica and Luca~Guido~Molinari}
\address{C.~A.~Mantica: I.I.S. Lagrange, Via L. Modignani 65, 
20161, Milano, Italy -- L.~G.~Molinari (corresponding author): Physics Department,
Universit\`a degli Studi di Milano and I.N.F.N. sez. Milano,
Via Celoria 16, 20133 Milano, Italy.}
\email{carloalberto.mantica@libero.it, luca.molinari@unimi.it}
\title[On the Weyl and Ricci tensors of GRW]
{On the Weyl and Ricci tensors\\ of generalized Robertson-Walker space-times}
\begin{abstract} 
We prove theorems about the Ricci and the Weyl tensors on 
generalized Robertson-Walker space-times of dimension $n\ge 3$. In particular,
we show that the concircular vector introduced by Chen decomposes the Ricci tensor 
as a perfect fluid term plus a term linear in the contracted Weyl tensor.   
The Weyl tensor is harmonic if and only if it is annihilated by Chen's vector, and any of the two
conditions is necessary and sufficient for the GRW space-time to be quasi-Einstein
(perfect fluid) manifold. Finally, the general structure of the Riemann tensor for Robertson-Walker 
space-times is given, in terms of Chen's vector. A GRW space-time in $n=4$ with 
null conformal divergence is a Robertson-Walker space-time.
\end{abstract}
\date{31 july 2016}
\maketitle
\section{Introduction}
The beautiful theorem by Chen (2014, \cite{BangYenChen}) gives a simple and 
covariant cha\-rac\-te\-risation of Generalized Robertson-Walker space-times (GRW).
It states that a Lorentzian manifold of dimension $n\ge 3$ is a GRW if and only if there 
exists a vector field $X_j$, which we name {\em Chen's vector},
such that $X^jX_j<0$ and
\begin{align}
\nabla_j X_k = \rho g_{jk}  \label{chen}
\end{align}
where $\rho $ is some scalar field. The equation is the defining property of a concircular vector, introduced by Fialkow \cite{Fialkow} 
(Yano gives the same name to a broader class of vectors \cite{Yano}). 
The existence of such a vector field is a 
necessary and sufficient condition for a local expression
of the metric with the warped form  
\begin{align}\label{warped}
  ds^2 = - dt^2 + q^2(t)g^*_{\mu\nu}(\vec x) dx^\mu dx^\nu 
 \end{align}
where $g^*_{\mu\nu}$ is the metric tensor of a Riemannian submanifold $M^*$ parametrised by $\vec x$.
The local form \eqref{warped} is actually how GRW manifolds were defined in 1995 by Alias et al. \cite{ARS}, and 
studied subsequently (see for example \cite{San98,Deszcz99,GutOl}). 

The strong property of Chen's vector upon differentiation allows for a determination of 
several interesting properties of GRW manifolds. The following two sections are devoted to the Ricci and to the Weyl tensor respectively. 
After showing that Chen's vector is an eigenvector of the Ricci tensor, the Ricci tensor is expressed as the sum of a perfect fluid term
and a term proportional to the Weyl tensor.\\ 
The main theorem states that the Weyl tensor is harmonic ($\nabla^m C_{jklm}=0$)  if and only if $C_{jklm} X^m=0$.
Any of the two conditions is necessary and sufficient for the
GRW manifold to be a quasi-Einstein (or perfect fluid) space-time,  i.e. for the Ricci tensor to have the form
$R_{ij} = A g_{ij}+ B u_iu_j$. \\
Part of these statements existed in the literature:  
in 1994 Gebarowski \cite{Gebarowski} proved that the fibers of a GRW space-time are Einstein 
(i.e. the Ricci tensor of the sub-manifold $M^*$ in the decomposition \eqref{warped} has the 
form $R^*_{\mu\nu} = R^*g_{\mu\nu}/(n-1)$)
if and only if $\nabla^m C_{jklm}=0$. Then, Sanchez (1999, \cite{Sanchez99}) stated that the fibers of a GRW space-time are Einstein 
if and only if the GRW manifold is quasi-Einstein (or perfect-fluid). Recently, Mantica et al. (2016, \cite{ManticaMolinariDe}) showed the converse: 
a perfect-fluid Lorentzian manifold with $\nabla^m C_{jklm}=0 $ is a GRW space-time (see also \cite{MSD}).\\
In this presentation, besides other results, we show that the same propositions hold with the algebraic condition $C_{jklm}X^m=0$ where $X$ is Chen's vector. \\
In the last section we consider conformally flat GRW space-times, i.e. Robertson-Walker space-times, and provide
the general form of the Riemann tensor. In par\-ti\-cu\-lar, the aforementioned results imply that in $n=4$ a conformally
harmonic GRW space-time is a Robertson-Walker space-time.

\section{The Ricci tensor}
\begin{thrm}
On a GRW manifold:\\
1) Chen's vector $X_j$ is an eigenvector of the Ricci tensor, 
\begin{align} 
R_{jm} X^m = \xi X_j , \qquad  \label{RicciX}
\end{align}
and the following holds for the eigenvalue:
\begin{align}
\xi = -  (n-1) \frac{X^k\nabla_k\rho}{X^2}, \quad \nabla_i\xi = \theta X_i \label{theta}
\end{align} 
where $\theta $ is a scalar field.\\
2) the Ricci tensor can be expressed in terms of the Weyl tensor, the curvature scalar $R$, the eigenvalue
$\xi$ and Chen's vector:
\begin{align}
R_{kl}=(n-2)C_{jklm}\frac{X^jX^m}{X^2} + \frac{R-\xi}{n-1}\left (g_{kl}-\frac{X_kX_l}{X^2}\right )+\xi \frac{X_kX_l}{X^2} ;\label{RicciWeyl}
\end{align}
\begin{proof}
By Chen's theorem, on a GRW manifold there is a vector field $X_j$ such that $\nabla_iX_j = \rho g_{ij}$. Then $[\nabla_i,\nabla_j] X_k = (\nabla_i \rho) g_{jk} -( \nabla_j\rho) g_{ik}$ i.e. $R_{ijk}{}^m X_m = (\nabla_i \rho) g_{jk} - (\nabla_j\rho) g_{ik}$. A contraction with $g^{jk}$ gives $R_{im}X^m  = -(n-1)\nabla_i \rho$, while
the contraction with $X^k$ gives $0=X_j\nabla_i\rho - X_i\nabla_j\rho $, then $X^2 \nabla_j \rho = X_j (X^i\nabla_i \rho)$ i.e. $\nabla_i\rho $ is proportional to $X_i$. Therefore \eqref{RicciX} follows.\\
Moreover:
\begin{align}
R_{ijk}{}^m X_m = -\frac{\xi}{n-1}( X_i g_{jk} - X_j g_{ik} ) \label{RieX}
\end{align}
The derivative of \eqref{RieX} is 
$$X^m\nabla_s R_{ijkm} + \rho R_{ijks}=-\frac{\nabla_s\xi}{n-1}(X_i g_{jk} - X_j g_{ik} ) -\frac{\rho \xi}{n-1}(g_{si} g_{jk} -  g_{ik} g_{js}), $$
the sum on cyclic permutations of indices $sij$ and the second Bianchi identity give:
$$ 0 = g_{jk} (X_i\nabla_s\xi -X_s\nabla_i\xi)+ g_{ik}(X_s\nabla_j\xi - X_j\nabla_s\xi) + g_{sk}(X_j\nabla_i\xi-X_i\nabla_j\xi). $$ 
The contraction with $g^{sk}$ finally gives $X_j\nabla_i\xi - X_i\nabla_j\xi =0$, with solution $\nabla_i\xi = \theta X_i$, thus proving
\eqref{theta}.\\
By definition, the Weyl tensor is
$$ C_{jklm} = R_{jklm} +\frac{1}{n-2}(g_{jm}R_{kl}-g_{km}R_{jl} + R_{jm}g_{kl} - R_{km}g_{jl}) - R\frac{ g_{jm}g_{kl}-g_{jl}g_{km}}{(n-1)(n-2)} $$
A contraction with $X^m$ and  \eqref{RieX} give:
\begin{align}
C_{jkl}{}^mX_m = \frac{\xi -R}{(n-1)(n-2)}(X_jg_{kl}-X_kg_{jl}) + \frac{1}{n-2}(X_j R_{kl}-X_k R_{jl}). \label{CX}
\end{align}
Another contraction with $X^j$ gives the Ricci tensor \eqref{RicciWeyl}.
\end{proof}
\end{thrm}
A covariant derivative of the eigenvalue equation \eqref{RicciX}, and use of \eqref{chen} and
$\nabla^k R_{kj} = \frac{1}{2}\nabla_j R$, where $R$ is the curvature scalar, gives a relation 
that will be important in the sequel: 
\begin{align}
\tfrac{1}{2}X^k\nabla_k R = n\rho\xi -\rho R + X^2\theta \label{crucial}
\end{align}

Multiplication of \eqref{RieX} by $X_l$ and summation on cyclic permutations of $ijl$ shows that Chen's vector
is ``Riemann compatible''\cite{RCT}: $X_iX^m R_{jlkm} + X_j X^m R_{likm} + X_l X^m R_{ijkm} =0 $.
In general, Riemann implies Weyl compatibility \cite{WCT}. In the present case (it can be checked with \eqref{CX}):
\begin{align}
X_iX^m C_{jklm} + X_j X^m C_{kilm} + X_k X^m C_{ijlm} =0 \label{weylcomp}
\end{align}

\section{The Weyl tensor}
We now focus on the Weyl tensor. It is useful to introduce the auxiliary symmetric trace-less tensor 
$${\sf C}_{jk}=C_{ajkb}\frac{X^aX^b}{X^2}$$
Note the properties $X^j {\sf C}_{jk}=0$ and $X^j \nabla_l {\sf C}_{jk}= - (\nabla_l X^j)
{\sf C}_{jk} = -\rho {\sf C}_{kl}$, that will be frequently used.\\
The contraction of \eqref{weylcomp} by $X^i$ or, the insertion of the expression of the Ricci tensor in \eqref{CX}, gives
\begin{prop}
The Weyl tensor of a GRW manifold satisfies the identity:
\begin{align}
C_{jklm}X^m = X_j {\sf C}_{kl} -X_k {\sf C}_{jl} \label{CXCXC}
\end{align}
\end{prop}
\noindent
It implies that $C_{jklm}X^m=0$ if and only if ${\sf C}_{kl}=0$.

The general expression for the covariant divergence of the Weyl tensor is:
\begin{align}
\nabla^m C_{jklm} = -\frac{n-3}{n-2}\left [ \nabla_j R_{kl} -\nabla_k R_{jl} 
-\frac {g_{kl} \nabla_j R - g_{jl} \nabla_k R}{2(n-1)} \right ]. \label{divC}
\end{align}
We look for an expression in terms of the contracted tensor ${\sf C}_{jk}$. The following covariant derivatives are evaluated with \eqref{chen} and $\nabla_j\xi = \theta X_j$: 
\begin{align}
 \nabla_j R_{kl} -\nabla_k R_{jl} &= (n-2)(\nabla_j {\sf C}_{kl} -\nabla_k {\sf C}_{jl}) \nonumber \\
&+\frac{1}{n-1} (n\rho\xi -\rho R+X^2\theta )
\left ( \frac{X_k}{X^2} g_{jl} - \frac{X_j}{X^2} g_{kl}\right) \nonumber \\
& + \frac{1}{n-1} \left [ \left( g_{kl} - \frac{X_kX_l}{X^2} \right ) \nabla_j R 
- \left( g_{jl}-\frac{X_jX_l}{X^2}\right ) \nabla_k R \right ] \nonumber \\
& =\frac{1}{n-1} \Big [  \left( g_{kl} - \frac{X_kX_l}{X^2} \right ) \nabla_j R 
- \left( g_{jl}-\frac{X_jX_l}{X^2}\right ) \nabla_k R \label{stepone}\\
&+\frac{1}{2}\left ( \frac{X_kX_s}{X^2} g_{jl} - \frac{X_jX_s}{X^2} g_{kl}\right)\nabla^s R\Big ] 
+ (n-2)(\nabla_j {\sf C}_{kl} -\nabla_k {\sf C}_{jl}) \nonumber
\end{align}
because of the identity \eqref{crucial}. Eq.\eqref{divC} becomes:
\begin{align}
& \nabla^m C_{jklm} = - (n-3) (\nabla_j {\sf C}_{kl} -\nabla_k {\sf C}_{jl} )  
+ \frac{n-3}{(n-1)(n-2)}  \frac{X_l}{X^2}  (X_k \nabla_j R - X_j \nabla_k R ) \label{steptwo}  \\ 
& -  \frac{n-3}{ 2(n-1)(n-2) } \left [ g_{kl} \left ( g_{jm} - \frac{X_j X_m}{X^2} \right ) \nabla^m R - 
  g_{jl} \left( g_{km} - \frac{X_kX_m}{X^2} \right ) \nabla^m R \right ] \nonumber
\end{align}
Its contraction with $g^{kl}$ gives
\begin{align}
\nabla^k {\sf C}_{jk} = \frac{n-3}{2(n-1)(n-2)} \left ( \nabla_j R  - \frac{X_jX^l}{X^2}\nabla_l R\right ). 
\label{CscalarR} 
\end{align}
\begin{lem}
\begin{align}
X^j \nabla_j {\sf C}_{kl} =-2 \rho {\sf C}_{kl} \label{auxaux}
\end{align}
\begin{proof}
The contraction of \eqref{steptwo} with $X^j$ is: 
\begin{align*}
X^j\nabla^m C_{jklm} = -(n-3) (X^j \nabla_j {\sf C}_{kl} +\rho {\sf C}_{kl}) -
\tfrac{(n-3)X_l}{2(n-1)(n-2)}\left (\nabla_k R -\frac{X_kX^j}{X^2} \nabla_j R\right )
\end{align*}
With the aid of \eqref{CscalarR} and with a permutation of indices, it becomes:
\begin{align}
\nabla^j (C_{jlkm} X^m)=  -(n-3) (X^j \nabla_j {\sf C}_{kl} +\rho {\sf C}_{kl}) -X_l \nabla^m {\sf C}_{km}
\end{align}
The left-hand-side of this equation is evaluated by means of \eqref{CXCXC}: $\nabla^j (C_{jlkm} X^m) = 
\nabla^j (X_j {\sf C}_{lk} -X_l {\sf C}_{jk})$ i.e.
\begin{align}
\nabla^j (C_{jlkm} X^m) = (n-1)\rho {\sf C}_{kl} +X^j\nabla_j {\sf C}_{kl} -X_l \nabla^j {\sf C}_{jk}
\end{align}
The two equations imply \eqref{auxaux}.
\end{proof}
\end{lem}
The following statement is important:
\begin{prop}
If $X^l\nabla^m C_{jklm}=0$ then:
\begin{gather}
\nabla_i R = X_i \frac{X^m\nabla_m}{X^2}R \label{R}\\
\nabla^m C_{jklm} = -(n-3) (\nabla_j {\sf C}_{kl} -\nabla_k {\sf C}_{jl})   \label{stepthree}
\end{gather}
\begin{proof}
Recall that $X^l\nabla_j {\sf C}_{kl}= -\rho {\sf C}_{jk}$. 
The contraction of \eqref{steptwo} with $X^l$ gives
\begin{align}
X^l\nabla^m C_{jklm} =\frac{n-3}{2(n-1)(n-2)}(X_k\nabla_j R-X_j\nabla_k R) 
\end{align}
If $X^l\nabla^m C_{jklm}=0$ then $X_k\nabla_j R = X_j\nabla_k R$, with solution \eqref{R}. 
Eq.\eqref{steptwo} greatly simplifies and reduces to \eqref{stepthree}.
\end{proof}
\end{prop}
Since $X^l\nabla^mC_{jklm} = \nabla^m(C_{jklm}X^l)$, the proposition holds in particular if $\nabla^mC_{jklm}=0$ or if $X^mC_{jklm}=0$.\\
We are ready to prove the main theorem:
\begin{thrm}\label{thrm_main}
On a GRW space-time with Chen vector $X_j$
\begin{align}
 C_{jklm}X^m=0\quad \Longleftrightarrow \quad \nabla^m C_{jklm}=0
 \end{align}
\begin{proof}
If $C_{jklm}X^m=0$ then ${\sf C}_{jk}=0$. The right-hand-side of \eqref{stepthree} is zero, and the Weyl tensor is harmonic.\\
If $\nabla^m C_{jklm}=0$ then $\nabla_j {\sf C}_{kl} -\nabla_k {\sf C}_{jl}=0$. In particular:
\begin{align}
 X^j \nabla_j {\sf C}_{kl} = -\rho {\sf C}_{kl} \label{aux0}
\end{align}
Because of \eqref{auxaux} it is ${\sf C}_{kl}=0$ i.e. $C_{jklm}X^m=0$. 
\end{proof}
\end{thrm}
\begin{prop}\label{prop_iff}
On a GRW space-time with Chen vector $X_k$, $C_{jklm}X^m =0$ if and only if 
\begin{align}
R_{jk}=\alpha g_{jk} +\beta \frac{X_jX_k}{X^2} \label{Ric_pf}
\end{align}
for suitable scalars $\alpha $ and $\beta $. 
\begin{proof}
If $C_{jklm}X^m=0$, then \eqref{RicciWeyl} gives $R_{jk}$ the perfect fluid form. If $R_{jk}=\alpha g_{jk}+
\beta \frac{X_jX_k}{X^2}$ then $R=n\alpha +\beta $ and $\xi = \alpha + \beta $, and the left hand side of 
\eqref{CX} is zero. 
\end{proof}
\end{prop}
Because of the main theorem \ref{thrm_main} we also have:
\begin{prop}\label{prop_iff2}
On a GRW space-time, it is $\nabla^m C_{jklm}=0 $ if and only if the Ricci tensor has the
structure \eqref{Ric_pf} (i.e. the manifold is quasi-Einstein).
\end{prop}
\section{Robertson-Walker space-times}
Robertson-Walker space-times are an important subclass of GRW space-times; they share the property
of being conformally flat, $C_{jklm}=0$. Let us then consider GRW space-times that are conformally flat.

The following theorem applies \cite{BV}: 
A GRW space-time is conformally flat if and only if the GRW manifold is the ordinary Robertson-Walker space-time
(or: if and only if the sub-manifold $M^*$ in the warped product \eqref{warped}
is a space of constant curvature).\\
With $C_{jklm}=0$ the Ricci tensor has the structure \eqref{Ric_pf} and,
in view of theorem \ref{thrm_main} and proposition \ref{prop_iff}, the Riemann tensor is largely determined: 
\begin{align} 
&R_{jklm} =  \frac{2\xi-R}{(n-1)(n-2)} (g_{kl}g_{jm}- g_{km}g_{jl} ) \label{Riemann}\\
&+\frac{ R-n\xi }{(n-1)(n-2) } \left [ g_{jm} \frac{X_kX_l}{X^2} -g_{km} \frac{X_jX_l}{X^2} +
g_{kl}\frac{X_jX_m}{X^2}  - g_{jl} \frac{X_k X_m}{X^2} \right ] \nonumber  \end{align}
where $X$ is Chen's vector, $R$ is the curvature scalar, $\xi $ is the eigenvalue of the Ricci tensor with
eigenvector $X$.\\  
Eq. \eqref{Riemann} is the general form of the Riemann tensor of a Robertson-Walker space-time. The
form characterises manifolds of quasi-constant curvature, introduced by Chen and Yano in 1972, 
\cite{chen72}.

In four-dimensions, the Weyl tensor on a pseudo-Riemannian manifold has a special property:
if $C_{jklm}u^m=0$, where $u^ku_k\neq 0$, then $u_i C_{jklm}+ u_j C_{kilm} + u_k C_{ijlm} =0$ 
(see \cite{LovRun} page 128). In particular, a contraction with $u^i$ gives $C_{jklm}=0$.
As a consequence we may state:
\begin{prop}
In $n=4$, a GRW manifold with $\nabla^m C_{jklm}=0$ is a Robertson-Walker space-time.
\begin{proof}
In a GRW the condition $\nabla^m C_{jklm}=0$ is equivalent to $C_{jklm}X^m=0$. Then, in $n=4$,
it is $C_{jklm}=0$.
\end{proof}
\end{prop}

\end{document}